# Diamagnetic Susceptibility of Interacting Electrons System

Adam Badra Cahaya[1*]

[1] Department of Physics, Faculty of Mathematics and Natural Sciences, Universitas Indonesia

Paramagnetism or diamagnetism of a material are shown by parallel or antiparallel directions, respectively, of the induced magnetization under the influence of external magnetic field. Theoretical study of paramagnetic susceptibility and diamagnetic susceptibility of non-interacting electrons system are well described by Pauli's spin susceptibility and Landau-Peierls' orbital susceptibility, respectively. For interacting electron systems, the paramagnetic susceptibility has been widely studied by using Hubbard's electron – electron interaction. However, due to its smaller value, the diamagnetic susceptibility has not been studied. To investigate the enhancement of an interacting electrons system, we study a generalized space and time-dependent orbital susceptibility of conduction electron with a repulsive Hubbard's electron – electron interaction. By using the retarded Green function expression of the space and time-dependent orbital susceptibility, we found that the orbital susceptibility is enhanced when Hubbard's electron – electron interaction is negative.

*Corresponding author: Tel:+62-82139244909 e-mail: adam@sci.ui.ac.id

# 1. Introduction

The magnetism of a metallic material, such as rare earth based metals, can arise from the spin angular momentum or orbital angular momentum due to electric current vortices [1]. Theoretical study of magnetism due to spin and orbital angular momentum of non-interacting electrons system are well described by Pauli's spin susceptibility and Landau-Peierls' orbital susceptibility, respectively [2], [3]. Hubbard interaction has been used for describing the spin-susceptibility $\chi_{int}^{spin}$ of the interacting electrons systems [4]

$$\chi^{spin} = \frac{\chi_0^{spin}}{1 - UN(E_F)}. \qquad (1)$$

Here $U$ is Hubbard parameter that characterizes the repulsive electron – electron interaction. $\chi_0^{spin} = \mu_B^2 N(E_F)$ is the spin-susceptibility of non-interacting electrons. $\mu_B$ is the Bohr magneton. The proportionality factor $\left(1 - UN(E_F)\right)^{-1}$ is often called Stoner enhancement, which depends on the Hubbard parameter $U$ and density of state at Fermi energy $N(E_F)$ [5]. Since orbital susceptibility is one third of spin susceptibility [6]

$$\chi_0^{orbital} = -\frac{1}{3}\chi_0^{spin}, \qquad (2)$$

modification of orbital susceptibility of the interacting electron systems is not well-discussed.

To investigate the enhancement of an interacting electrons system, we study a generalized space and time-dependent orbital susceptibility of conduction electron and its relation to current response function in Section 2. In Section 3, we show how

Hubbard interaction modify the current response function of conduction electron. By describing the interacting electron system using Hubbard interaction, we discuss the Stoner enhancement of orbital susceptibility in Section 4. The findings are summarized in Section 5.

## 2. Orbital susceptibility formalism using current response function

To describe the susceptibility of an interacting electrons systems, a Hubbard's interaction Hamiltonian $H_1$ describing an effective repulsive electron – electron interaction can be added to the free electronic Hamiltonian $H_0$ [4], [5], [7]

$$H = H_0 + H_1 = \sum_{k\sigma} E_k a^\dagger_{k,\sigma} a_{k,\sigma} + \sum_{kpq} U a^\dagger_{p+q,\uparrow} a_{p,\uparrow} a^\dagger_{k-q,\downarrow} a_{k,\downarrow}. \qquad (3)$$

Here $E_k = \hbar^2 k^2/2m$, $a^\dagger_{k,\sigma}$ and $a_{k,\sigma}$ are energy, creation and annihilation operators of conduction electron with wavenumber $\mathbf{k}$, respectively. $U$ is Hubbard parameter that characterizes the repulsive interaction between conduction electrons [8].

The magnetization $\mathbf{M}$ of the system can arise from the collective movement of conduction electrons due to external magnetic field $\mathbf{B}$. The interaction energy can be written in the following integral

$$H_{\text{ext}} = -\int d^3 r\, \mathbf{M} \cdot \mathbf{B} = -\int d^3 r\, \mathbf{j} \cdot \mathbf{A}. \qquad (4)$$

From the relation $\mathbf{B} = \nabla \times \mathbf{A}$, one can see that the orbital magnetization $\mathbf{M}$ is related to the electric current vortex

$$\mathbf{M} = \nabla \times \mathbf{j} \qquad (5)$$

One can see that $\mathbf{M} = -\frac{1}{2}\mathbf{r} \times \mathbf{j}$ fulfilled Eq. (5). Since the electric current $\mathbf{j} = -e\mathbf{p}/m$ is directly proportional to the collective momentum $\mathbf{p}$, one can see that the orbital magnetization $\mathbf{M}$ arise from the orbital angular momentum of conduction electrons

$$\mathbf{M} = -\frac{1}{2}\mathbf{r} \times \mathbf{j} = \frac{e}{2m}\mathbf{r} \times \mathbf{p} = \gamma_L \mathbf{L} \qquad (6)$$

Here $\gamma_L = e/2m$ is the Zeeman orbital gyromagnetic ratio.

The electric current induced by external magnetic field can be obtained using Kubo formula [9]

$$j_a(\mathbf{r},t) = \frac{1}{i\hbar}\int_{-\infty}^{t} dt' \langle [j_a(\mathbf{r},t), H_{ext}(t')] \rangle$$

$$= \int d^3r' \int_{-\infty}^{\infty} dt' R_{ab}(\mathbf{r}-\mathbf{r}',t-t') A_b(\mathbf{r}',t'). \qquad (7)$$

Here the current – current response function $R(\mathbf{r}-\mathbf{r}',t-t')$ can be written in the following retarded Green function expression

$$R_{ab}(\mathbf{r}-\mathbf{r}',t-t') = \frac{i}{\hbar}\theta(t-t')\langle [j_a(\mathbf{r},t), j_b(\mathbf{r}',t')] \rangle, \qquad (8)$$

$\theta(t-t')$ is the Heaviside step function. Later, it will be shown to be isotropic $R_{ab} = \delta_{ab}R$. Using symmetric Gauge $\mathbf{A} = \frac{1}{2}\mathbf{B} \times \mathbf{r}$ [10], [11], one can determine the orbital susceptibility tensor from the current response function

$$M_a = \chi_{ab}^{orbital} B_b = R\left(\frac{1}{4}\mathbf{r} \times (\mathbf{r} \times \mathbf{B})\right)_a$$

$$\chi_{ab}^{orbital} = R(r_a r_b - r^2 \delta_{ab}) \qquad (9)$$

The effect of $U$ on $R$ of interacting system is discussed in Section 3.

## 3. Mean field approximation of current – current response function

Using the second quantization expression of electric current

$$\mathbf{j}(\mathbf{r}) = -\frac{e\hbar}{m}\sum_{\mathbf{pq}\sigma} e^{i\mathbf{q}\cdot\mathbf{r}}\left(\mathbf{p}+\frac{\mathbf{q}}{2}\right) a^{\dagger}_{\mathbf{p+q},\sigma} a_{\mathbf{p},\sigma}, \qquad (10)$$

one can see that the Fourier transform of $R_{ab}(\mathbf{r})$ to momentum space is

$$R_{ab}(\mathbf{q}) = -\frac{ie}{m}\theta(t)\sum_{\mathbf{p}\sigma}\left(\mathbf{p}+\frac{\mathbf{q}}{2}\right)_a \langle[a^{\dagger}_{\mathbf{p+q},\sigma} a_{\mathbf{p},\sigma}, j_{0b}(\mathbf{0},0)]\rangle$$

$$\equiv \sum_{\mathbf{p}\sigma}\left(\mathbf{p}+\frac{\mathbf{q}}{2}\right)_a S_{b\sigma}(\mathbf{p},\mathbf{q}). \qquad (11)$$

$S_{b\sigma}$ can be evaluated using its time derivation

$$\frac{\partial}{\partial t}S_{b\sigma}(\mathbf{p},\mathbf{q}) = -\frac{ie}{m}\delta(t)\langle[a^{\dagger}_{\mathbf{p+q},\sigma} a_{\mathbf{p},\sigma}, j_{0b}(\mathbf{0},0)]\rangle$$

$$-\frac{ie}{m}\theta(t)\langle\left[\frac{1}{i\hbar}[a^{\dagger}_{\mathbf{p+q},\sigma} a_{\mathbf{p},\sigma}, H], j_{0b}(\mathbf{0},0)\right]\rangle. \qquad (12)$$

The first term can be approximated using random phase approximation $a^{\dagger}_{\mathbf{p},\alpha} a_{\mathbf{q},\beta} \simeq \delta_{\mathbf{pq}}\delta_{\alpha\beta} f_{\mathbf{p},\alpha}$

$$\langle[a^{\dagger}_{\mathbf{p+q},\sigma} a_{\mathbf{p},\sigma}, j_{0b}(\mathbf{0},0)]\rangle \simeq \frac{e\hbar}{m}\left(\mathbf{p}+\frac{\mathbf{q}}{2}\right)_b (f_{\mathbf{p},\sigma} - f_{\mathbf{p+q},\sigma}) \qquad (13)$$

The second term can be evaluated using mean field approximation

$$a^{\dagger}_{\mathbf{p}\alpha} a_{\mathbf{q}\alpha} a^{\dagger}_{\mathbf{k}\beta} a_{l\beta} \simeq f_{kl} a^{\dagger}_{\mathbf{p}\alpha} a_{\mathbf{q}\alpha} + f_{\mathbf{pq}} a^{\dagger}_{\mathbf{k}\beta} a_{l\beta} \qquad (14)$$

Such that

$$\langle[[a^{\dagger}_{\mathbf{p+q},\sigma} a_{\mathbf{p},\sigma}, H], j_{0b}(\mathbf{0},0)]\rangle = (E_{\mathbf{p}} - E_{\mathbf{p+q}})\langle[a^{\dagger}_{\mathbf{p+q},\sigma} a_{\mathbf{p},\sigma}, j_{0b}(\mathbf{0},0)]\rangle$$

$$+U(f_{\mathbf{p},\sigma} - f_{\mathbf{p+q},\sigma})\sum_{\mathbf{k}}\langle[a^{\dagger}_{\mathbf{k+q},\sigma} a_{\mathbf{k},\sigma}, j_{0b}(\mathbf{0},0)]\rangle \qquad (15)$$

Using Eq. (11) one can see that

$$i\hbar\frac{\partial}{\partial t}S_{b\sigma}(\mathbf{p},\mathbf{q},t) = \left(\frac{e\hbar}{m}\right)^2 \delta(t)\left(\mathbf{p}+\frac{\mathbf{q}}{2}\right)_b (f_{\mathbf{p},\sigma} - f_{\mathbf{p+q},\sigma}) + (E_{\mathbf{p}} - E_{\mathbf{p+q}})S_{b\sigma}(\mathbf{p},\mathbf{q},t)$$

$$+U(f_{\mathbf{p+q},\sigma} - f_{\mathbf{p},\sigma})\sum_{\mathbf{k}} S_{b\sigma}(\mathbf{k},\mathbf{q},t) \qquad (16)$$

In frequency domain,

$$S_{b\sigma}(\mathbf{p},\mathbf{q},\omega) = \frac{\left(\frac{e\hbar}{m}\right)^2 \left(\mathbf{p}+\frac{\mathbf{q}}{2}\right)_b (f_{\mathbf{p},\sigma} - f_{\mathbf{p+q},\sigma})}{(E_{\mathbf{p+q}} - E_{\mathbf{p}} + \hbar\omega)} + U\frac{(f_{\mathbf{p+q},\sigma} - f_{\mathbf{p},\sigma})}{(E_{\mathbf{p+q}} - E_{\mathbf{p}} + \hbar\omega)}\sum_{\mathbf{k}\sigma} S_{b\sigma}(\mathbf{k},\mathbf{q},t) \qquad (17)$$

Taking sum over $\mathbf{p}$

$$\sum_{\mathbf{p}} S_{b\sigma}(\mathbf{p},\mathbf{q},t) = \sum_{\mathbf{p}} \frac{\left(\frac{e\hbar}{m}\right)^2 \left(\mathbf{p}+\frac{\mathbf{q}}{2}\right)_b (f_{\mathbf{p},\sigma}-f_{\mathbf{p+q},\sigma})}{(E_{\mathbf{p+q}}-E_{\mathbf{p}}+\hbar\omega)} + \sum_{\mathbf{k}\sigma} S_{b\sigma}(\mathbf{k},\mathbf{q},t) U \sum_{\mathbf{p}} \frac{(f_{\mathbf{p+q},\sigma}-f_{\mathbf{p},\sigma})}{(E_{\mathbf{p+q}}-E_{\mathbf{p}}+\hbar\omega)}$$
(18)

Since the summation in the last term only depends on $\mathbf{q}$ and $\omega$,

$$\sum_{\mathbf{p}} S_{b\sigma}(\mathbf{p},\mathbf{q},t) = \sum_{\mathbf{k}} S_{b\sigma}(\mathbf{k},\mathbf{q},t) = \zeta(\mathbf{q},\omega) \sum_{\mathbf{p}} \frac{\left(\frac{e\hbar}{m}\right)^2 \left(\mathbf{p}+\frac{\mathbf{q}}{2}\right)_b (f_{\mathbf{p},\sigma}-f_{\mathbf{p+q},\sigma})}{(E_{\mathbf{p+q}}-E_{\mathbf{p}}+\hbar\omega)}$$
(19)

where the enhancement factor

$$\zeta = \left(1 - U \sum_{\mathbf{p}} \frac{(f_{\mathbf{p+q},\sigma}-f_{\mathbf{p},\sigma})}{(E_{\mathbf{p+q}}-E_{\mathbf{p}}+\hbar\omega)}\right)^{-1}$$
(20)

Because of that, we can set

$$S_{b\sigma}(\mathbf{p},\mathbf{q},t) = \zeta \left(\frac{e\hbar}{m}\right)^2 \left(\mathbf{p}+\frac{\mathbf{q}}{2}\right)_b \frac{(f_{\mathbf{p},\sigma}-f_{\mathbf{p+q},\sigma})}{(E_{\mathbf{p+q}}-E_{\mathbf{p}}+\hbar\omega)}.$$
(21)

Substituting Eq. (21) back to Eq. (11) one can show that $R_{ab}(\mathbf{q}) = \delta_{ab} R(q)$ is isotropic

$$R_{ab}(\mathbf{q}) = \zeta \left(\frac{e\hbar}{m}\right)^2 \sum_{\mathbf{p}\sigma} \left(\mathbf{p}+\frac{\mathbf{q}}{2}\right)_a \left(\mathbf{p}+\frac{\mathbf{q}}{2}\right)_b \frac{(f_{\mathbf{p},\sigma}-f_{\mathbf{p+q},\sigma})}{(E_{\mathbf{p+q}}-E_{\mathbf{p}}+\hbar\omega)}$$

$$= \delta_{ab} \zeta \frac{e^2 k_F^3}{8\pi^2 m} \left(\frac{5}{3} - \left(\frac{q}{2k_F}\right)^2 + \frac{(1-(q/2k_F)^2)^2}{q/k_F} \ln\left|\frac{q+2k_F}{q-2k_F}\right|\right) \equiv \delta_{ab} R(q) \quad (22)$$

## 4. Orbital susceptibility of interacting electrons system

Orbital susceptibility tensor for static limit $\omega \to 0$ can be found by substituting Eq. (22) to Eq. (9)

$$\chi_{ab}^{\text{orbital}}(q) = -\left(\frac{\partial}{\partial q_a}\frac{\partial}{\partial q_a} - \delta_{ab}\nabla_q^2\right) R(q)$$

$$= -\frac{e^2 k_F}{2^9 \pi^2 m} \frac{\zeta(q)}{(q/k_F)^3} \left(\delta_{ab}\left[16\frac{q}{k_F} + 36\left(\frac{q}{k_F}\right)^3 - \left(16 - 8\left(\frac{q}{k_F}\right)^2 + 9\left(\frac{q}{k_F}\right)^4\right)\ln\left|\frac{q+2k_F}{q-2k_F}\right|\right] + \frac{q_a q_b}{q k_F}\left[-48\frac{q}{k_F} - 12\left(\frac{q}{k_F}\right)^3 + \left(48 + 8\left(\frac{q}{k_F}\right)^2 + 3\left(\frac{q}{k_F}\right)^4\right)\ln\left|\frac{q+2k_F}{q-2k_F}\right|\right]\right)$$
(23)

Fig. 1 illustrate the dependency on $q$ of the isotropic and anisotropic terms of $\chi_{ab}^{\text{orbital}}(q)$. Negative values in $q \to 0$ limit indicates that the orbital susceptibility is a diamagnetic response. Change of sign near $q = 2k_F$ is due to divergence in $q = 2k_F$, which is a typical signature of metallic systems [2].

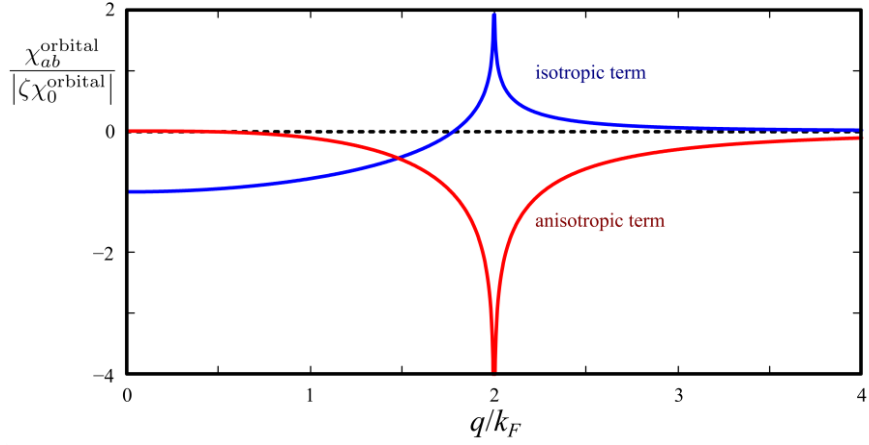

Figure 1. Orbital susceptibility of interacting conduction electrons system $\chi_{ab}^{orbital} = \delta_{ab}\chi_{isotropic}^{orbital} + \hat{q}_a\hat{q}_b\chi_{anisotropic}^{orbital}$. The enhancement factor $\zeta = 1/(1 + UN(E_F))$ indicates that positive $U$ reduces $\chi_{ab}^{orbital}$. $U$ is Hubbard parameter that characterize interacting electron system. $N(E_F)$ is the density of states at Fermi energy.

The enhancement factor can be evaluated using

$$\sum_p \frac{(f_{p+q,\sigma}-f_{p,\sigma})}{(E_{p+q}-E_p+\hbar\omega)} = -N(E_F)\left(\frac{1}{2} + \frac{1-\left(\frac{q}{2k_F}\right)^2}{\frac{2q}{k_F}}\ln\left|\frac{q+2k_F}{q-2k_F}\right|\right) \quad (24)$$

Which in small $q \ll k_F$ limit, is similar to the enhancement of spin susceptibility in Eq. (1)

$$\lim_{q \ll k_F} \zeta(q) = \frac{1}{1+UN(E_F)} \quad (25)$$

But with opposite sign of $UN(E_F)$. Because of that, we can conclude that the electron – electron interaction reduces the orbital susceptibility

$$\chi_{ab}^{orbital} = \delta_{ab} \frac{\chi_0^{orbital}}{1+UN(E_F)}. \qquad (26)$$

Here $\chi_0^{orbital}$ is the orbital susceptibility for non-interacting electrons system. For uniform magnetic fields, we can focus on small $q \ll k_F$ limit. In this case, the leading order of the orbital susceptibility tensor is isotropic

$$\lim_{q \ll k_F} \chi^{orbital} = -\frac{\mu_B^2 N(E_F)}{3} \frac{1}{1+UN(E_F)}. \qquad (27)$$

For $U = 0$, we recover $\chi_0^{orbital} = -\chi_0^{spin}/3$. For general values of $U$, the dependence of $\chi^{orbital}$ can be represented in Fig. 2. Fig. 2 also illustrates the total magnetic susceptibility for an interacting electron system, which can be found by adding the paramagnetic and diamagnetic susceptibilities in Eqs. (1) and (27), respectively. One can see that the interacting electron system becomes a superdiamagnet, diamagnet, paramagnet and ferromagnet for $UN(E_F)<-1$, $-1<UN(E_F)<-0.5$, $-0.5<UN(E_F)<1$ and $UN(E_F)>1$, respectively.

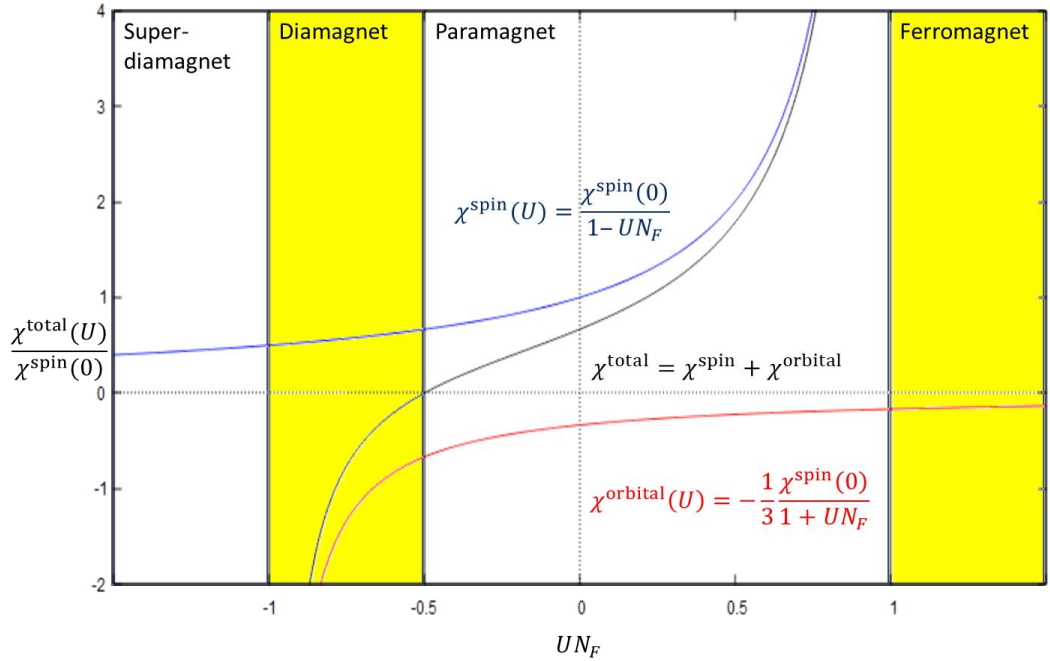

Figure 2. Dependency of spin and orbital susceptibilities of interacting conduction electrons can be characterized by $UN(E_F)$ parameter.

## 5. Summary and conclusion

An interacting electron system is characterized by repulsive electron – electron interaction. We represent the repulsive interaction using Hubbard's interaction Hamiltonian that is characterized by parameter *U*. The orbital magnetization M arises from the electric current vortex induced by external magnetic field. The current vortex corresponds with the orbital angular momentum of the conduction electron. By using symmetric Gauge, the orbital susceptibility tensor can be related to the current – current response function. $(1+UN(E_F))^{-1}$ factor indicates that the repulsive nature of electrons in interacting system reduces its orbital magnetic moment.